\documentclass[11pt,twoside]{article}

\usepackage{asp2006}
\usepackage{graphicx}

\begin{document}
\title{AGN Observations in the GeV/TeV Energy Range with the MAGIC Telescope}
\author{R. M. Wagner for the MAGIC Collaboration}
\affil{Max-Planck-Institut f\"{u}r Physik, F\"{o}hringer Ring 6, D-80805 M\"{u}nchen, Germany}

\begin{abstract}
MAGIC currently is the largest imaging atmospheric Cerenkov
telescope world-wide. Since 2004, $\gamma$-ray emission from
several active galactic nuclei in the GeV/TeV energy range has
been detected, some of which were newly discovered as very-high
energy $\gamma$-ray sources. The $\gamma$-rays are assumed to
originate from particle acceleration processes in the AGN jets. We
give an overview of the AGN observed and detected by MAGIC,
discuss spectral and temporal properties of these and show physics
implications of some selected observations.
\end{abstract}

\section{Introduction}
The study of very high energy (VHE, $E\geq 100$ GeV) $\gamma$-ray
emission from active galactic nuclei (AGN) is one of the major
goals of ground-based $\gamma$-ray astronomy. The sensitivity of
the current imaging air Cerenkov telescopes (IACT) enables
phenomenological studies of the physics inside the relativistic
jets in blazars, and in particular advances in understanding both
the origin of the VHE $\gamma$-rays as well as the relations
between photons of different energies (from radio to VHE).

Except for the radio galaxy M87, all 17 currently known VHE
gamma-ray emitting AGNs \citep{Wagner2,Wagner3} are
blazars,\footnote{See {\tt
http://www.mppmu.mpg.de/$\sim$rwagner/sources/} for an up-to-date
source list.} which are characterized by a close orientation of
the jets with the line of sight. Their spectra are dominated by
non-thermal emission that consists of two distinct broad
components. While the low energy bump, located at optical to X-ray
energies, is uni\-so\-nous\-ly explained by synchrotron emission
of electrons, the origin of the high-energy part of the spectrum
is still debated. Leptonic models ascribe it to inverse Compton
processes that either up-scatter synchrotron photons, or to
external photons that originate from the accretion disk, from
nearby massive stars, or are reflected into the jet by surrounding
material. In hadronic models, interactions of a highly
relativistic jet outflow with ambient matter, proton-induced
cascades, or synchrotron radiation off protons, are responsible
for the high energy photons.

Another defining property of blazars is the high variability of
their emission ranging from radio to $\gamma$-rays. For VHE
$\gamma$-ray blazars, correlations between X-ray and $\gamma$-ray
emission have been found on time scales ranging from several 10
minutes to days \citep[e.g., ][]{fossati04}, although the relation
has proven to be rather complicated.

Here we present selected results for MAGIC blazar observations of
Mkn~421, Mkn~501, 1ES\,2344+514, PG 1553+113, and BL~Lacert\ae.

\section{The MAGIC Telescope}
MAGIC \citep{MAGIC-commissioning,cortina} is located on the Canary
Island of La Palma (2200~m a.s.l.) and is currently the largest
IACT. Its energy range spans from 60~GeV (trigger threshold at
small zenith angles) up to tens of TeV. MAGIC has a sensitivity of
$\sim$~2.5\% of the Crab Nebula flux in 50 observation hours. Its
energy resolution is about 30\% above 100 GeV and about 25\% from
200 GeV onwards. The MAGIC standard analysis chain is described
e.g. by \citet{Crab_MAGIC}. Observations during moderate moonshine
enable a substantially extended duty cycle \citep{jrico}, which is
particularly important for blazar observations. Parallel optical
observations are performed with the
KVA 35~cm telescope.

The MAGIC AGN observation program encompasses over 500 hours/year.
Besides participating in various multiwavelength campaigns
\citep[e.g., ][]{masaaki} and many {\it target of opportunity}
observation agreements \citep[e.g., ][]{danielexpliclyintext,
ackerman}, MAGIC is cooperating with other Cerenkov telescopes as
to extend the dynamical range and/or the time coverage of
observations \citep{hessmagic} and monitors some well-known bright
blazars on a regular basis \citep{monitoring}. Apart from having
observed previously known blazars, MAGIC has also newly discovered
a couple of blazars in VHE $\gamma$-rays, as 1ES\,1218+30.4
\citep{1218}, Mkn 180 \citep{magic180}, PG 1553+113
\citep{magic1553}, and recently BL~Lacert\ae\,\citep{bllac}.

\section{Markarian 421}
Mkn~421 is the closest ($z = 0.030$) known TeV blazar and, along
with Mkn~501, also the best studied one. It has shown variations
larger than one order of magnitude and occasional flux doubling
times as short as 15 minutes~\citep{gaidos}. Variations in the
hardness of the TeV $\gamma$-ray spectrum during flares were
reported by several groups \citep[e.g., ][]{krenn, 2155}.
Simultaneous observations in the X-ray and VHE bands showed a
significant correlation of the respective fluxes
\citep{krawczynski421}.

Mkn~421 was observed with MAGIC during 19 nights for a total of
25.6 hours, with observation times per night ranging from 30
minutes up to 4 hours. It was found to be in a medium flux state
ranging from 0.5 to 2 Crab units above 200~GeV.
Significant variations of up to an overall
factor of four and up to a factor two between successive nights
were found.
A clear correlation between the $2-10$~keV X-ray count rate,
obtained from the public RXTE All-Sky Monitor data,\footnote{These data are available at \texttt{http://xte.mit.edu/}.} and the VHE
$\gamma$-ray flux can be described well by both a linear fit
(correlation coefficient $ r = 0.64^{+0.15}_{-0.22}$) and a
parabolic fit (both forced to go through the origin).

\begin{figure}
    \hfill
    \begin{minipage}[t]{.42\textwidth}
      \begin{center}
                \includegraphics[width=\textwidth]{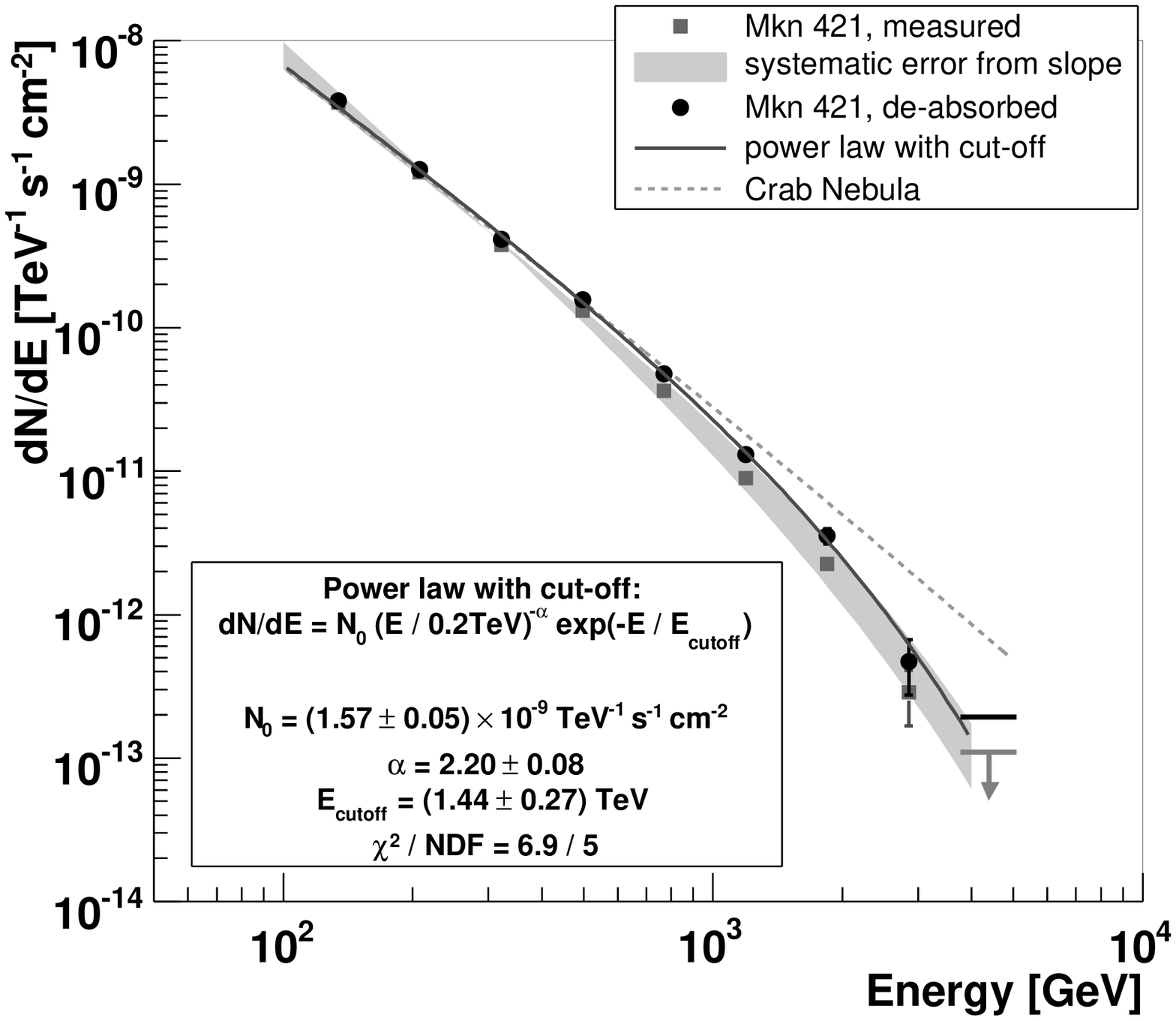}
   \caption{The measured (grey squares) and intrinsic (black points) spectrum
         of Mkn~421. Solid line: power-law + cutoff fit to the intrinsic spectrum.}
        \label{fig:421:3}
      \end{center}
    \end{minipage}
    \hfill
        \begin{minipage}[t]{.48\textwidth}
      \begin{center}
    \includegraphics[width=.95\textwidth]{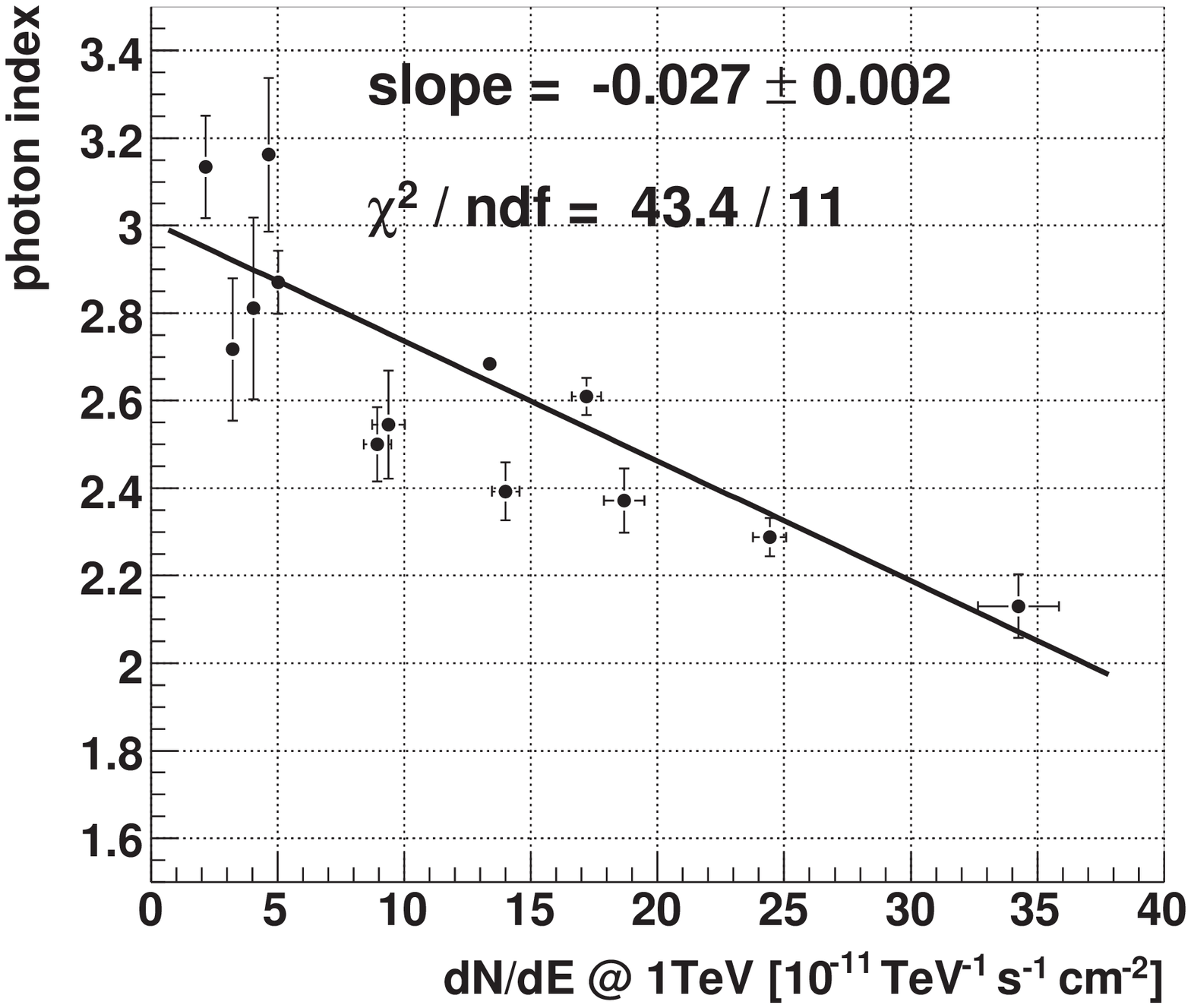}
   \caption{Correlation between spectral slope and differential flux for various VHE observations at 1 TeV for Mkn 421 \citep[for references, see][]{magic421}.}
        \label{fig:421:4}
      \end{center}
    \end{minipage}
    \hfill
  \end{figure}

The energy spectrum of Mkn 421, for the first time measured down
to $E=100$~GeV, is shown in Fig.~\ref{fig:421:3} along with the
reconstructed intrinsic spectrum. The latter has been obtained by
correcting for $\gamma\gamma$ absorption effects caused by
extragalactic background light (EBL; Hauser \& Dwek 2001), using
the recent EBL model of \citet{primack}. The intrinsic spectrum is
clearly curved, implying that the curvature in the measured
spectrum has an intrinsic origin rather than being caused by EBL
absorption. Both a power-law+exponential cutoff fit and a curved
power-law fit indicate a flattening of the spectrum towards
100~GeV. The spectrum is hardening with an increasing flux level
(Fig.~\ref{fig:421:4}). The analysis is described in detail in
\citet{magic421}.

\section{Markarian 501}
Mkn 501 ($z=0.034$) is known to be a strong and variable VHE
$\gamma$-ray emitter. During a flare in 1997 Mkn~501 showed strong
variability on timescales of 0.5 days, and the integral flux
reached 10 times the flux of the Crab nebula above $1$~TeV
\citep[e.g., ][]{hegra501}. While a correlation between spectral
hardness and $\gamma$-ray flux was found, the position of the
inverse Compton peak could not be detected.

\begin{figure}
    \hfill
    \begin{minipage}[t]{.58\textwidth}
      \begin{center}
        \includegraphics[width=\textwidth]{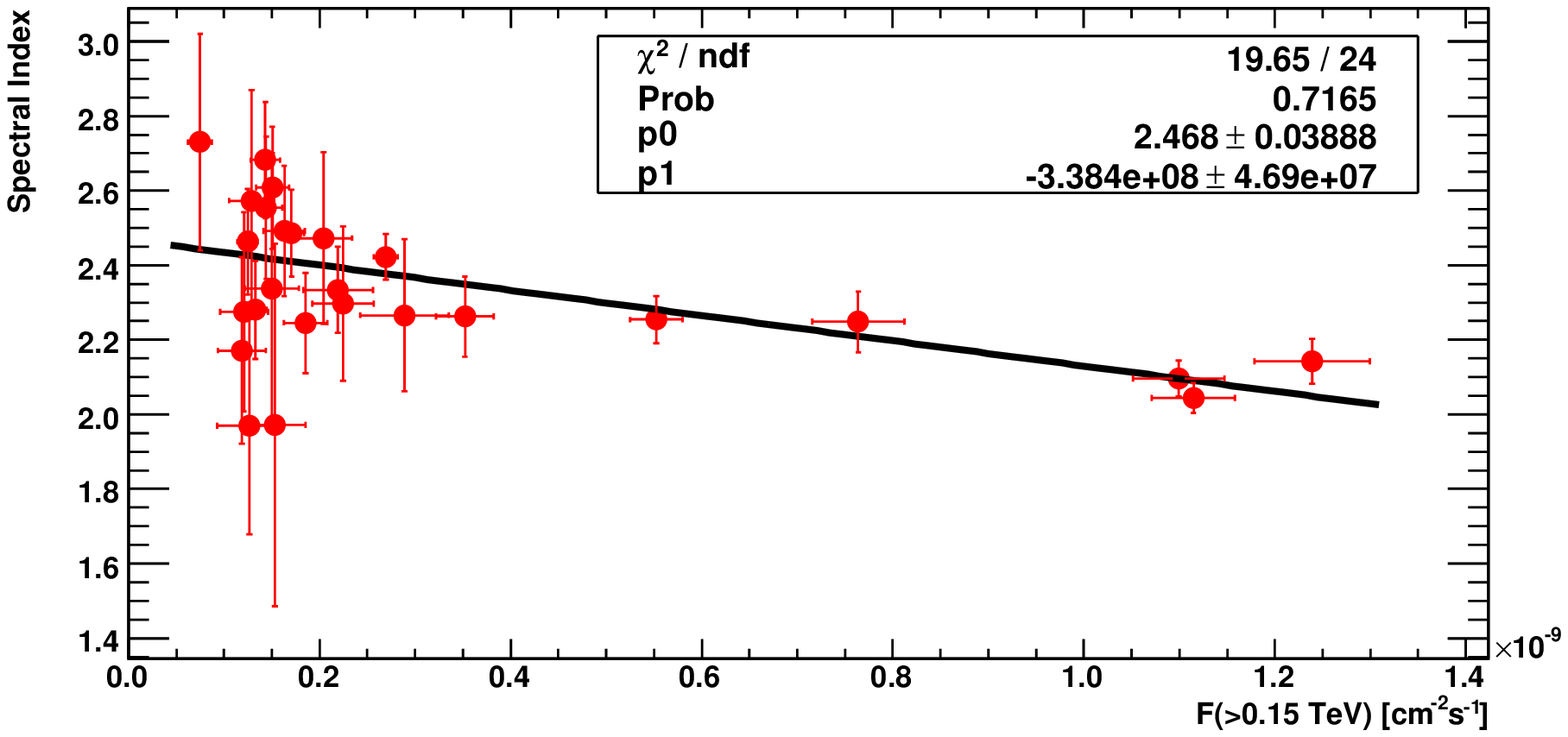}
   \caption{Correlation of spectral hardness and VHE $\gamma$-ray flux above 200 GeV. Each data point represents the average over one night, except for the four highest VHE flux points (two points per night).}
        \label{fig:501:4}
      \end{center}
    \end{minipage}
    \hfill
        \begin{minipage}[t]{.40\textwidth}
      \begin{center}
        \includegraphics[width=\textwidth]{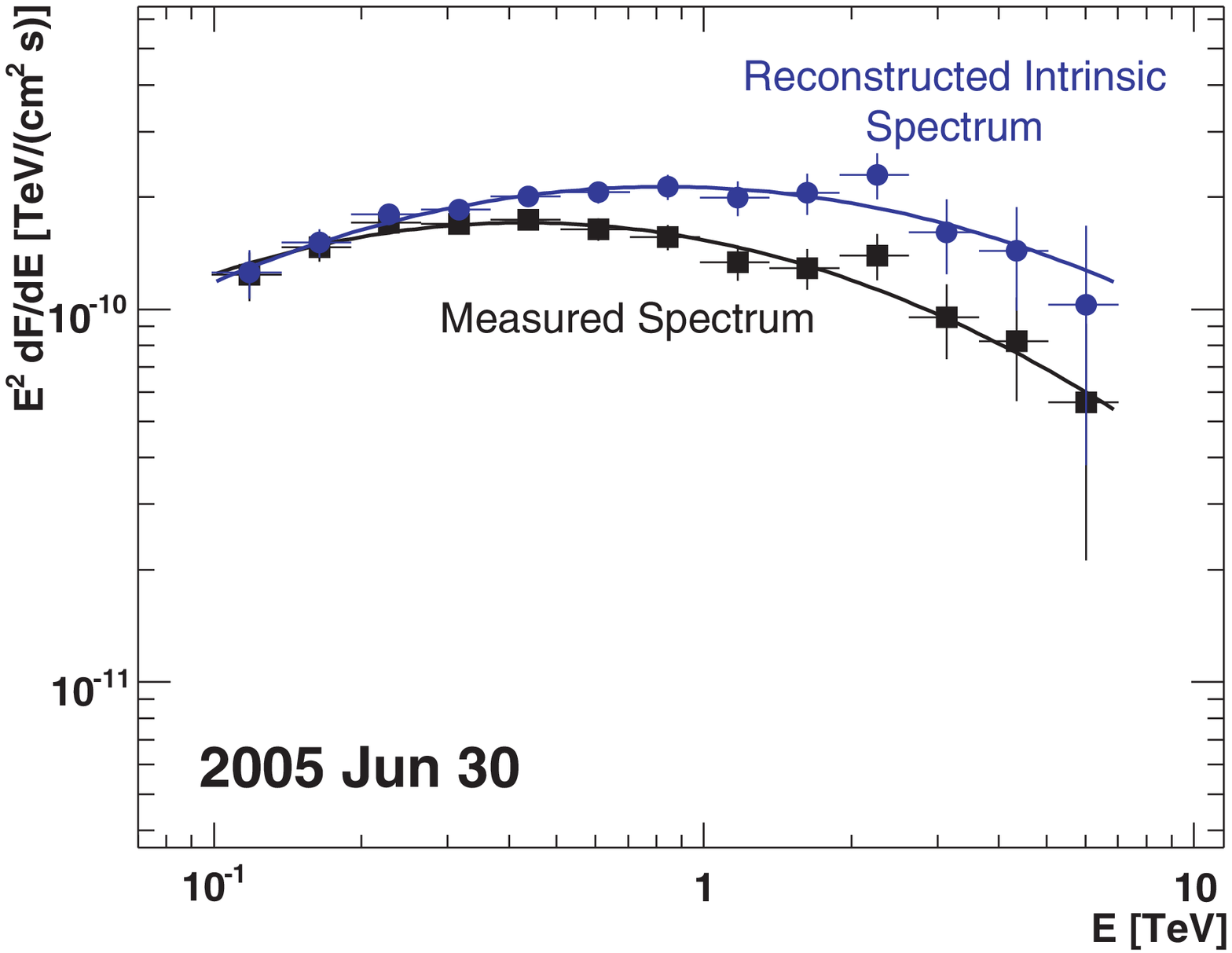}
   \caption{Measured (lower curve) and reconstructed intrinsic (upper curve) spectrum of Mkn 501 during the night of 2005 June 30.}
        \label{fig:501:5}
      \end{center}
    \end{minipage}
    \hfill
  \end{figure}

MAGIC observed Mkn~501 for 24 nights during six weeks in summer
2005. About 18~hours out of the total 30 observation hours were
performed in the presence of (moderate) moonshine. During most of
the observations the source was in a rather low flux state,
corresponding to 30\% of the Crab nebula flux at $E\geq200$~GeV.
In two nights (one with moon present), however, Mkn 501 was
reaching up to four times the Crab nebula flux. A correlation
between spectral hardness and VHE $\gamma$-ray flux was found also
for Mkn 501 (Fig.~\ref{fig:501:4}). For the flare nights, a clear
peak in the spectrum is seen (e.g., Fig.~\ref{fig:501:5}). In
leptonic models it can be identified with the ``Inverse Compton
peak'', which now for the first time is seen unambiguously in VHE
range for blazars.

\begin{figure}
        \includegraphics[width=\textwidth]{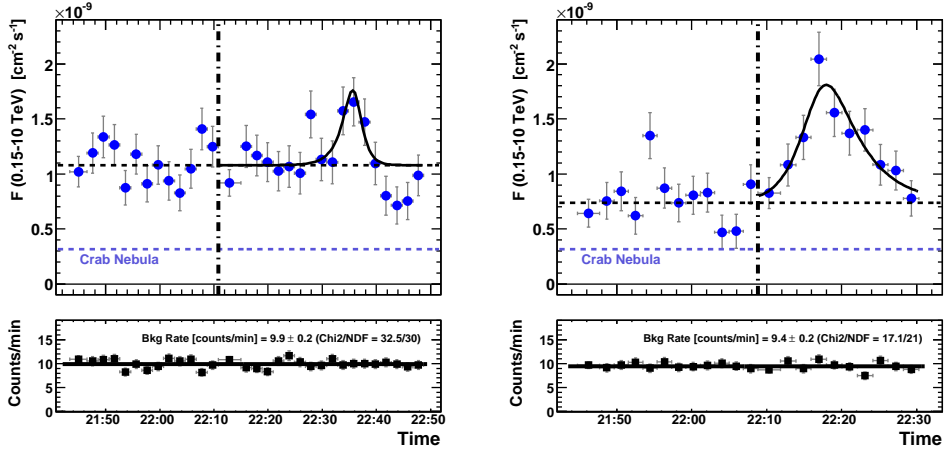}
   \caption{Light curves (2 min bin width) for the nights of 2005 June 30 (top left) and 2005 July 9 (top right). The lower plots show the corresponding background rates,
   which were found to be constant.}
        \label{fig:501:2}
  \end{figure}

\begin{figure}
  \begin{center}
        \includegraphics[width=.47\textwidth]{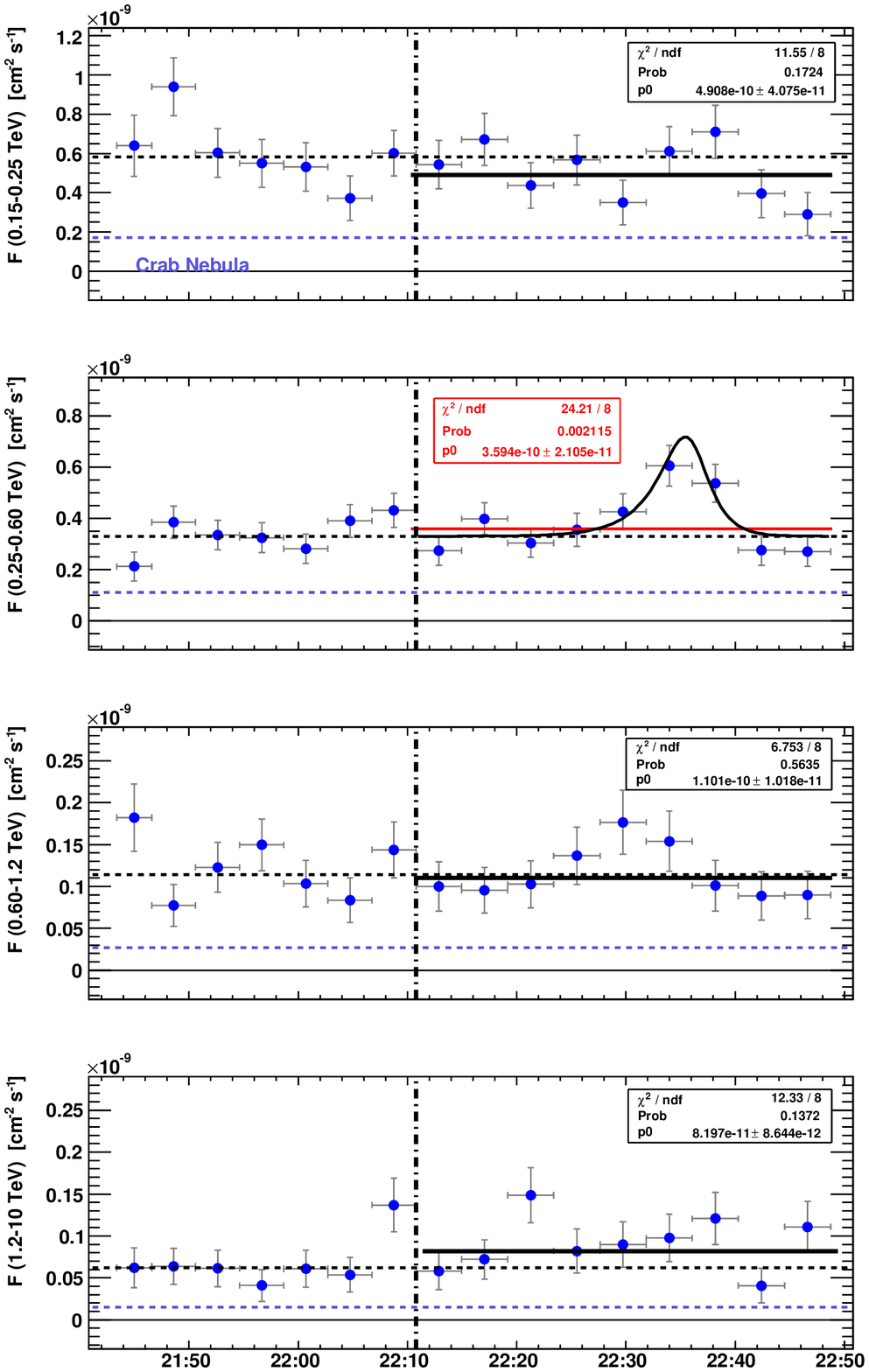}
        \includegraphics[width=.47\textwidth]{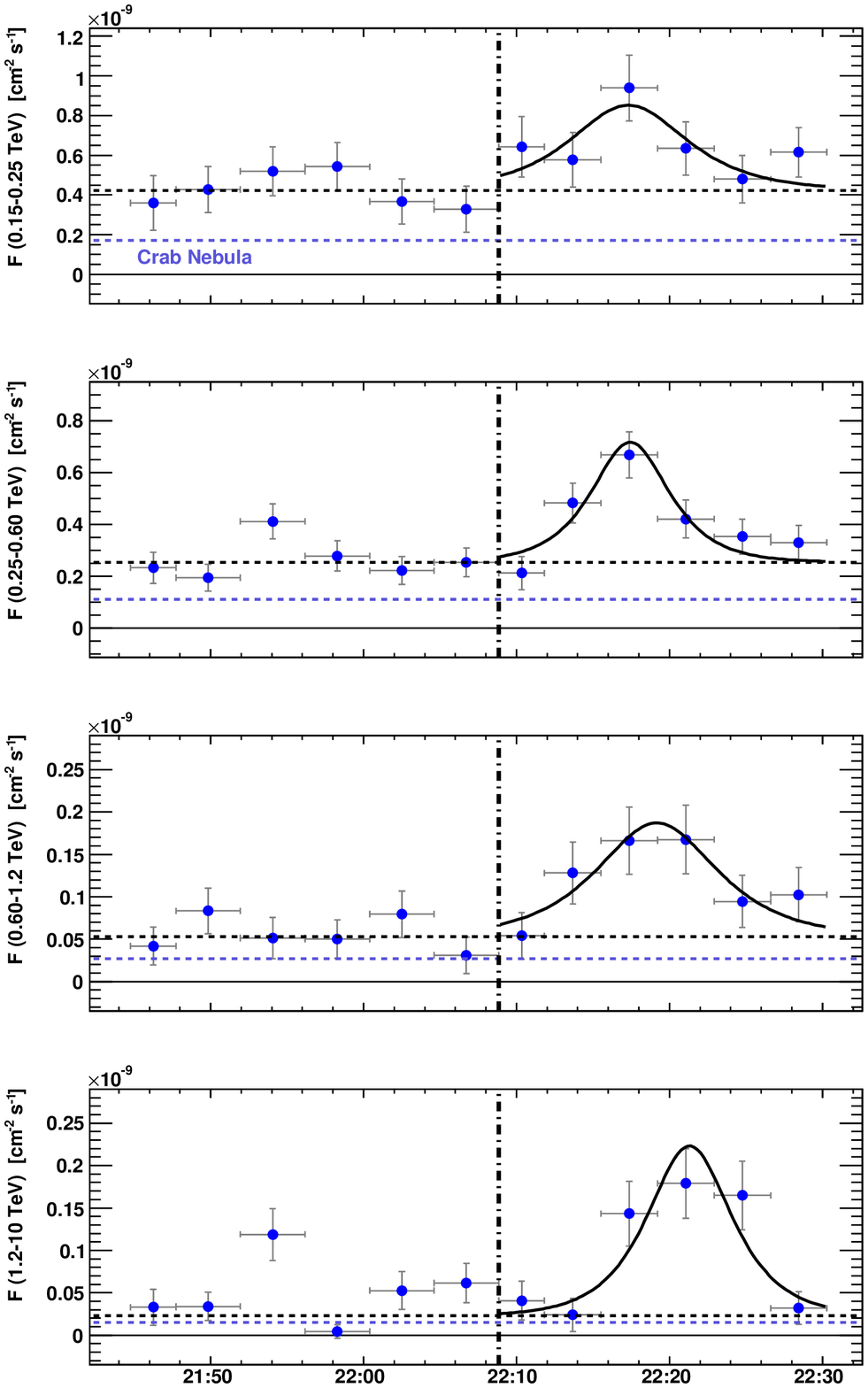}
   \caption{Evolution of the flares of 2005 June 30 (left column) and 2005 July 9 (right column) in distinct energy bins.}
        \label{fig:501:3}
          \end{center}
  \end{figure}

A closer look at the two flare nights reveals rapid flux changes
with doubling times as short as 3 minutes or less
(Fig.~\ref{fig:501:2}), while constant background rates assure
that no instrumental effects are responsible for these.
Fig.~\ref{fig:501:3} shows corresponding light curves in
different, distinct energy bands between 150 GeV and 10 TeV.
Interestingly the flares in the two nights behave differently:
While the 2005 June 30 flare is only visible in 250 GeV to 1.2
TeV, the 2005 July 09 flare is apparent in all energy bands, and a
marginal time delay towards higher energies is visible. For the
first time, short ($\approx$~20 min) flares with a resolved time
structure can now be subjected to detailed studies of particle
acceleration and cooling timescales. The analysis and more results
are shown in \cite{alb501}.

\section{1ES\,2344+514}
\label{sec:3}

1ES\,2344+514 ($z=0.044$) was detected in VHE $\gamma$-rays in
1995/6 on the $5.8\,\sigma$ level \citep{Catanese}. Over 1/3 of
the excess was recorded during one night in December 1995 with
$F_\gamma(\geq350\mathrm{GeV}) =$~64\% of the Crab nebula flux.
The $\gamma$-ray emission was confirmed on a $4.4\sigma$
significance level by \citet{54AGN} in observations in 1997-2002.

In 2005, MAGIC saw a clear $11\sigma$ excess from the direction of
1ES\,2344+514. The measured flux corresponds to 10\% of the Crab
nebula flux ($E>200$~GeV). For the first time a diurnal
measurement of the emission of this blazar was possible, yielding
no strong evidence for significant variability on times scales of
days and thus giving first insights to the day-scale properties of
a low blazar emission state. The differential energy spectrum can
be fitted by a simple power law with a photon index of
$\alpha=-2.95 \pm 0.12$. The derived spectrum is softer than the
one reported by the Whipple collaboration during the flare in 1995
\citep{schroedter}. Details on the analysis are given in
\citet{alb2344}.

\section{PG\,1553+113}
\begin{figure}
    \hfill
    \begin{minipage}[t]{.42\textwidth}
      \begin{center}
        \includegraphics[width=\textwidth]{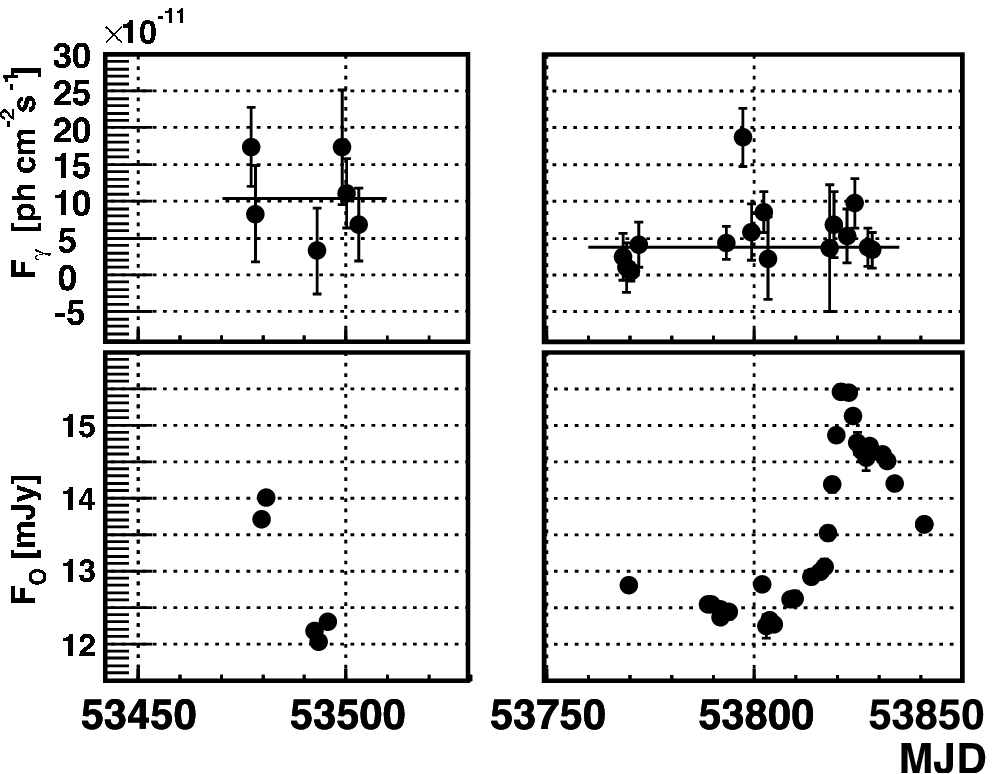}
   \caption{$E\geq180$~GeV and $R$-band light curves for PG\,1553+113.}
        \label{fig:1553:2}
      \end{center}
    \end{minipage}
    \hfill
        \begin{minipage}[t]{.48\textwidth}
      \begin{center}
        \includegraphics[width=\textwidth]{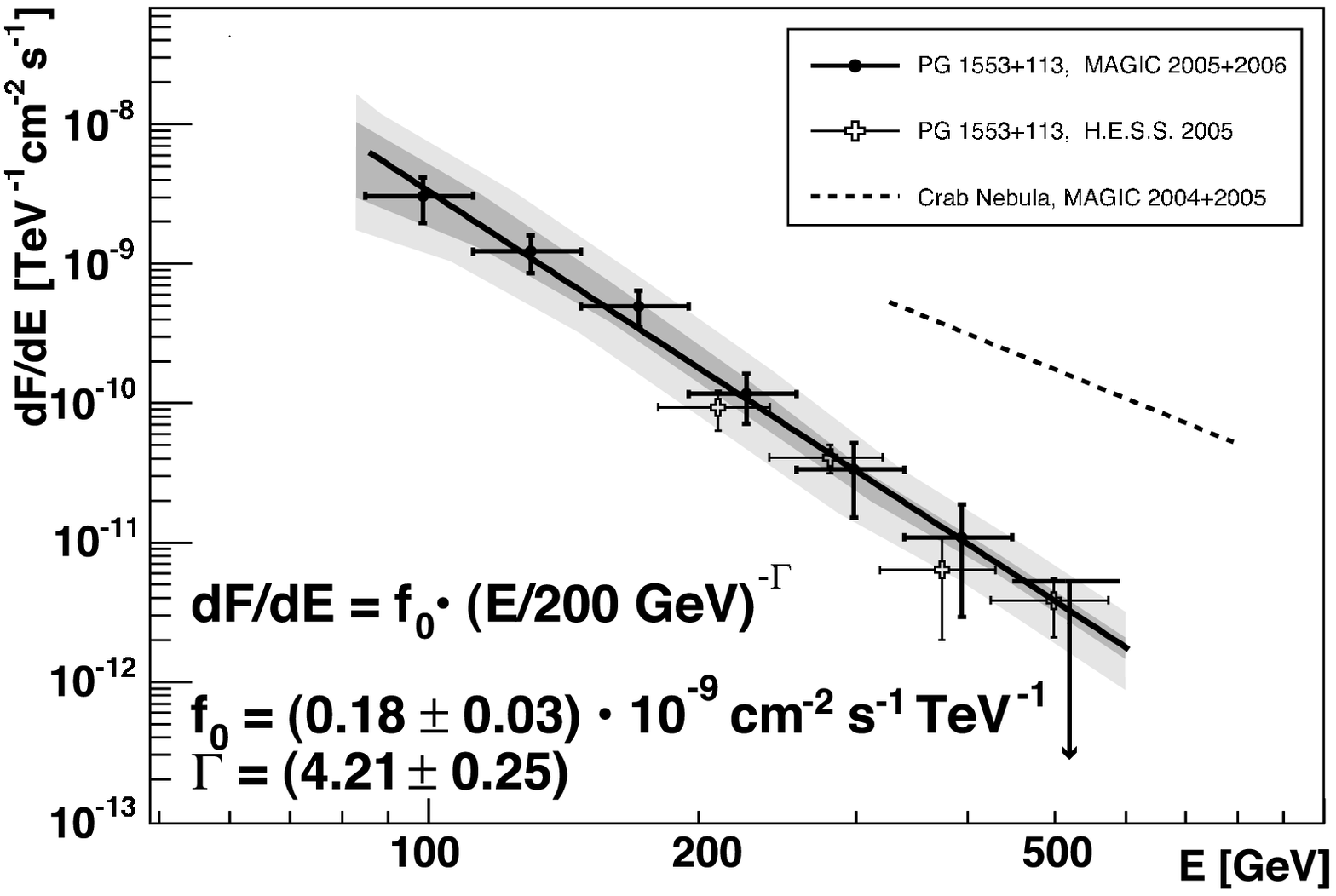}
   \caption{Differential energy spectrum for PG\,1553+113.}
        \label{fig:1553:3}
      \end{center}
    \end{minipage}
    \hfill
  \end{figure}

Based on its SED properties, PG 1553+113 was one of the most
promising candidates for VHE $\gamma$-ray emission
\citep{CostGhis}. The blazar was observed in 2005/6, yielding a
$\gamma$-ray signal at a significance level of $8.8\,\sigma$.
There is no evidence for short term variability on a time scale of
days, but a significant change by a factor of three in the flux
level from 2005 to 2006 was found. The VHE $\gamma$-ray flux was
not found to correlate with a substantial brightening of the
source in the optical $R$-band around MJD 53820
(Fig.~\ref{fig:1553:2}). The differential energy spectrum for PG
1553+113 is well described by a pure power law with a photon index
of $\alpha=-4.2\pm0.4$, the steepest photon index observed so far
for blazars (Fig.~\ref{fig:1553:3}). The signal detected by MAGIC
confirms a tentative signal seen by H.E.S.S. at a higher energy
threshold with data taken at about the same time as MAGIC in 2005
\citep{hess1553}. The source, therefore, can now be considered as
a firmly established VHE $\gamma$-ray emitter. Details of the
analysis are given in \citet{magic1553}.

The distance of PG~1553+113 is currently unknown, despite several
attempts of determining its redshift \citep[Direct measurements:
$z>0.09$, ][]{z_new}. The objects shows no emission lines, which
probably implies a close alignment of the jet with the line of
sight, or a jet outshining the galaxy core. Using the observed
spectrum of PG 1553+113 and assuming a minimum possible density of
the evolving EBL \citep[lower limit in ][]{kneiske}, an upper
limit of $z<0.74$ can be inferred when requiring the intrinsic
photon index $\Gamma < 1.5$. Other works employ SED fits
\citep{hess1553,magic1553}, the structure of the intrinsic
spectrum \citep{danflo}, or blazar ensemble arguments
\citep{Wagner2} to determine redshift limits.

\section{BL\,Lacert\ae}
BL\,Lacert\ae~($z=0.069$) is the historical prototype of the
BL~Lac class of AGNs. Unlike for the up to recently known ``VHE
blazars'' with their synchrotron peaks in the X-ray domain, the
peak of its synchrotron emission is located in the sub-millimeter
to optical band ($\nu_\textrm{peak}=2.2 \times 10^{14}$~Hz).

22.2 hours of MAGIC observations of BL Lacert\ae\ from August to
December 2005 resulted in a $5.1\sigma$ VHE $\gamma$-ray signal.
Above 200 GeV, an integral flux of $(0.6\pm0.2)\times10^{-11}~{\rm
cm}^{-2}~{\rm s}^{-1}$ was measured, corresponding to
approximately 3\% of the Crab nebula flux. No significant evidence
of flux variability was found. Optical and radio data shows
enhanced flux levels during the observations; While the VHE
$\gamma$-ray flux shows similar variations, no correlation can be
claimed at this point.

The differential spectrum between 150 and 900 GeV was found to be
rather steep, with a photon index of $\alpha=-3.6\pm0.5$. In
contrast to a 1997 flare of BL\,Lacert\ae, which necessitated
external photon fields to explain the VHE $\gamma$ emission
\citep{Rav02}, the MAGIC observations can be accommodated in a
simple SSC model. The detection is described in \citet{bllac}.

\end{document}